\documentclass[12pt,eqsecnum]{revtex4}
\begin{document}

\title {\Large Off-shell nilpotent finite BRST/anti-BRST transformations}

\author{ Sudhaker Upadhyay \footnote {e-mail address: sudhakerupadhyay@gmail.com}}
\author{ Sumit Kumar Rai \footnote{e-mail address: sumitssc@gmail.com}}
\author{ Bhabani Prasad Mandal \footnote{e-mail address:
\ \ bhabani.mandal@gmail.com ,\ \ bhabani@bhu.ac.in  }}

\affiliation { Department of Physics,\\
Banaras Hindu University,\\
Varanasi-221005, INDIA. \\
}

\begin{abstract}
We develop the off-shell nilpotent finite field dependent BRST transformations  and show 
that for different choices of the finite field dependent parameter these connect the generating 
functionals corresponding to different effective theories. We also construct both on-shell and off-shell finite field 
dependent anti-BRST tranformations for Yang Mills theories and
 show that these transformations  play the similar role in connecting different generating 
functionals of different effective theories. Analogous to the finite field dependent BRST
 transformations, the non-trivial Jacobians of the path integral measure which arise due to 
the finite field dependent anti-BRST transformations  are responsible for the new results.
 We consider several explicit examples in each case to demonstrate the results.
\end{abstract}
\maketitle

\section{\large Introduction}
The Becchi, Rouet, Stora and Tyutin (BRST) symmetry is a fundamental tool for the study of 
gauge theories. This symmetry guarantees the quantization, renormalizability, unitarity and 
other aspects of Yang-Mills (YM) 
theories \cite{brst,iv,ht,wei}. 
BRST symmetry is characterized by a continuous  
global parameter which is 
anticommuting in nature.
One of the most important 
characteristics of the BRST symmetry 
is nilpotency. The 
usual BRST transformation uses equation of motion to 
achieve nilpotency which is referred as on-shell nilpotency.
In order to make it nilpotent without using equations of 
motion, Nakanishi and Lautrup \cite{naoj} 
 introduced a new field, having no independent 
dynamics called as auxiliary field. 
In auxiliary field formulation BRST transformation  
becomes off-shell nilpotent.

Joglekar and Mandal have shown that the usual infinitesimal, global BRST transformation can be 
 integrated out to construct the finite field dependent BRST (FFBRST) transformation   
\cite{jm}. The parameter in such transformation is anticommuting, finite in nature, depends on the fields, and does not 
depend on space-time explicitly. FFBRST transformations are also the symmetry of 
the effective theories and maintain the on-shell nilpotency property.
FFBRST transformations can connect the 
generating functionals of two different effective field theories with suitable choice of  
the finite field dependent parameter \cite{jm}.
For example, these transformations can be used to connect the Faddeev-Popov (FP)
effective action in Lorentz gauge with a gauge parameter $\lambda $ to
(i) the most general BRST/anti-BRST symmetric action in Lorentz gauge \cite{jm}
, (ii) the FP effective action  in axial gauge \cite{sdj0,sdj2,sdj3,sdj4,sd}, (iii)  the 
FP effective action  in Coulomb gauge \cite{cou},
(iv) FP effective action  with another distinct gauge parameter 
$\lambda^\prime $ \cite{jm} and (v) the FP effective action in quadratic gauge \cite{jm}.
The FFBRST transformation can also be used to connect the generating functionals 
corresponding to different solutions of the quantum master equation
 in field/antifield formulation \cite{srbpm}.      
  The choice of the finite parameter  is crucial in connecting different
effective gauge theories by means of the FFBRST transformation. The path integral measure in the 
expression for generating functional is not invariant under FFBRST transformation. The 
non-trivial Jacobian of such FFBRST transformation is the source for the new results.

FFBRST transformations have found many applications \cite{sdj0,sdj2,sdj3,sdj4,sd,cou,sdj,rb,sdj1,etc} in the 
study of gauge theories. A correct prescription for the poles in the gauge field propagators in 
non-covariant gauges have been derived by connecting effective theories in covariant gauges to 
the theories in non-covariant
 gauges by using FFBRST transformation \cite{sdj1,cou}. The divergent energy integrals in the Coulomb gauge are 
regularized by modifying the time like propagator using FFBRST transformation \cite{cou}. The FFBRST transformation which is discussed so far, in literature is only on-shell nilpotent.

In this work, we formulate the FFBRST transformations in auxiliary field formulation to make 
these transformations off-shell nilpotent. We consider several examples to demonstrate that 
the FFBRST transformations in auxiliary field formulation also lead to similar results in connecting the different generating functionals. FP 
effective action remains invariant under a different form of BRST transformations where the 
roles of ghost and anti-ghost fields are interchanged, these BRST like transformations are known as 
anti-BRST transformations. Anti-BRST transformation does not play as fundamental role as BRST symmetry itself but it is a useful tool in geometrical description \cite{boto}  of BRST transformation, in the investigation of perturbative renormalization of Yang Mills models 
\cite{gaba}. We further construct the finite field dependent anti-BRST (FF anti-BRST) transformation analogous to 
the FFBRST transformation. By considering several similar choices of finite field dependent parameter in 
FF anti-BRST transformation we show that FF anti-BRST transformation also 
connects the generating functionals corresponding to different effective theories. Finally, 
 we consider 
the formulation of FF anti-BRST transformation in auxiliary field formulation also to make it off-shell nilpotent.

Now, we briefly mention the plan of the paper. We start with a small introduction to the FFBRST 
transformation in Sec. II and discuss the formulation of the FFBRST transformations in auxiliary field method 
with several examples in Sec. III. In Sec. IV, we construct FF anti-BRST transformation and consider few 
examples with different choices of finite parameter. In section V, we discuss the FF anti-BRST in 
auxiliary field formulation. We summarize the  results in Sec. VI.
\section{\large Preliminary : FFBRST}
Let us now briefly review the FFBRST formulation of pure gauge theories \cite
{jm,srbpm,sdj0,cou,sdj,rb,sdj1,etc}.
 FFBRST transformations are
obtained by integrating the infinitesimal  (field dependent ) BRST
transformations  \cite{jm}. In this method all the fields are functions
of some parameter, $ \kappa : 0\le \kappa \le 1$. For a generic field $ \phi (x, \kappa),\
\phi(x, \kappa =0 ) = \phi(x) $ is the initial field and
 $ \phi(x, \kappa=1) = \phi ^\prime
(x)$ is the transformed field. Then the infinitesimal field dependent BRST transformations are
defined as
\begin{equation}
\frac{ d}{d \kappa}\phi(x, \kappa ) = \delta _{BRST }\ \phi(x, \kappa )\
\Theta ^ \prime [\phi(x,\kappa )],
\label{ibr}
\end{equation}
where $\Theta ^\prime d \kappa $ is an infinitesimal field dependent parameter.
It has been shown \cite{jm} by integrating these equations from $ \kappa=0$ to $\kappa=1$
that the $\phi^\prime ( x) $ are related to $\phi(x)
$ by the FFBRST transformations
\begin{equation}
\phi^\prime (x) = \phi(x) + \delta _{BRST} \ \phi(x)\ \Theta [\phi(x)],
\label{fbrs}
\end{equation}
where $ \Theta [\phi(x)] $ is obtained from $\Theta ^\prime [\phi(x)] $
through the relation
\begin{equation}
\Theta [\phi(x)] = \Theta ^\prime [\phi(x)] \frac{ \exp f[\phi(x)]
-1}{f[\phi(x)]},
\label{80}
\end{equation}
and $f$ is given by $ f= \sum_i \frac{ \delta \Theta ^\prime (x)}{\delta
\phi_i(x)} \delta _{BRST}\ \phi_i(x). $
These transformations are nilpotent and symmetry of the FP effective action. 
However, 
Jacobian of the path integral measure changes the generating functional correponding to FP 
effective theory to 
the generating functional for a different effective theory.
               
The meaning of these field transformations is as follows. We consider
the vacuum expectation value of a gauge invariant functional $G[A]$
in some effective theory,
\begin{equation}
<< G[A]>> \equiv \int {\cal D} \phi \ G[A] \exp(iS_{eff}[\phi]),
\label{90}
\end{equation}
where
\begin{equation}
S_{eff} = S_0 + S_{gf}+S_g. 
\label{sf}
\end{equation}
Here, $S_0$  is the pure YM action
\begin{equation}
S_0= \int d^4x \left [-\frac{1}{4}F^{\alpha\mu\nu}F_{\mu\nu}^\alpha\right ],
 \label{s0}
\end{equation} and the gauge fixing and ghost part of the effective action in Lorentz gauge are
given as
\begin{eqnarray}
S_{gf} &=& -\frac{1}{2\lambda}\int d^4x (\partial \cdot{A^\alpha})^2 \nonumber,\\
S_g &=& -\int d^4x\left [ \bar{c}^\alpha\partial^\mu
D^{\alpha\beta}_\mu c^\beta \right ].
\label{s12}
\end{eqnarray}
The covariant derivative is defined as $ D^{ab}_\mu [A] \equiv \delta ^{ab}\partial
_\mu + g f^{abc}A^c_\mu $. 
                                                                                
Now we perform the FFBRST transformation $\phi\rightarrow \phi^\prime
 $ given by Eq. (\ref{fbrs}). Then we have 
\begin{equation}
<<G[A]>>=  <<G[A^\prime ]>> = \int {\cal D} \phi ^\prime
J[\phi^\prime ] G[A^\prime ] \exp(iS^F_{eff}[\phi^\prime ]),
\label{def}
\end{equation}
on account of BRST invariance of $S_{eff}$ and the  gauge invariance of $G[A]$.
Here $J[\phi^\prime ]$ is the Jacobian associated with FFBRST transformation and is defined
as
\begin{equation}
{\cal D} \phi = {\cal D} \phi^\prime J[\phi^\prime ].
\label{jac}
\end{equation}
Note that unlike the usual infinitesimal BRST transformation, the Jacobian
for FFBRST is not unity. In fact, this non-trivial Jacobian
is the source of the new results in this formulation. 
As shown in Ref. \cite{jm} for the special case $G[A]=1$, the Jacobian
$ J[\phi^\prime ]$ can always be replaced by
$ \exp(iS_1[\phi^\prime ]$),  where $S_1(\phi^\prime)$ is some {\it local}
functional of the fields and can be added to the action,
\begin{equation}
S_{eff}[\phi^\prime ] +S_1[\phi^\prime ] = S_{eff}^\prime
[\phi^\prime].
\label{s1}
\end{equation}
Thus the FFBRST transformations change the effective action of the theory.
\section{\large OFF-SHELL NILPOTENT FFBRST }
In this section, we intend to 
generalize the FFBRST 
formulation in an auxiliary field formulation. We only mention the necessary steps of the FFBRST 
formulation in presence of  auxiliary field \cite{srbpm}. For simplicity, we consider the case 
of pure 
YM theory described by the effective action in Lorentz gauge 
\begin{equation}
 S^L_{eff}= \int d^4x\left [-\frac{1}{4}F^{\alpha \mu \nu }F^{\alpha}_{\mu \nu } +\frac{\lambda 
}{2}(B^\alpha) ^2 - B^{\alpha}
\partial \cdot{A^\alpha }-\bar{c}^\alpha \partial^{\mu}D^{
\alpha \beta}_\mu  c^\beta \right ]. \label{seff}
\end{equation} 
Following the procedure outlined in section II, it is straightforward to construct FFBRST transformations under which the above $S^L_{eff}$ remain 
invariant.
These transformations are as follows:
\begin{eqnarray}
A^{\alpha}_\mu &\rightarrow & A^\alpha_\mu + D_\mu^{\alpha\beta}c^\beta \  \Theta
(A,c,\bar{c},B),\nonumber \\
c^\alpha &\rightarrow & c^\alpha -\frac{g}{2}f^{\alpha\beta\gamma}c^\beta c^\gamma \ \Theta(A,
 c, \bar{c},B), \nonumber \\
\bar{c}^\alpha &\rightarrow & \bar{c}^\alpha +B^\alpha \ \Theta(A,c,\bar{c},B),\nonumber \\
B^\alpha &\rightarrow & B^\alpha.
\label{ffb}
\end{eqnarray}
The finite parameter, $\Theta (A,c,\bar c,B)$  also depends on the auxiliary field B.
However, a non-trivial modification arises in the calculation of Jacobian for this FFBRST in 
an auxiliary field formulation. The Jacobian can now be defined as
\begin{eqnarray}
DA(x)Dc(x)D\bar{c}(x)DB(x) &=&J(x,k) DA(x,k)Dc(x,k)D\bar{c}(x,k)DB(x,k) \nonumber \\
              &=& J(k+dk)DA(k+dk)Dc(k+dk)D\bar{c}(k+dk)DB(k+dk).\nonumber\\
                                                                 \label{jac1}
\end{eqnarray}
The transformation from $\phi(k)$ to $\phi(k+dk)$ is an infinitesimal one and one has, for 
its Jacobian
\begin{equation}
\frac{J(k)}{J(k+dk)}=\sum_{\phi }\pm\frac{\delta{\phi(x,k+dk)}}{\delta{\phi(x,k)}},
\end{equation}
where $\sum_{\phi }$ sums over all the fields in the measure 
$A_\mu^\alpha,c^\alpha,\bar{c}^\alpha,B^\alpha $ and the $ \pm $ sign refers to the cases of fields $\phi$ being
bosonic or fermionic in nature.
We evaluate the right hand side as
\begin{equation}
\int d^4 x \sum_{\alpha}\left [ \sum_{\mu}\frac{\delta A_\mu^\alpha(x,k+dk)}{\delta A_
\mu^\alpha(x,k)} - \frac{\delta c^\alpha(x,k+dk)}{\delta c^\alpha(x,k)} - \frac{\delta\bar{c}^
\alpha(x,k+dk)}{\delta\bar{c}^\alpha(x,k)} + \frac{\delta B^\alpha(x,k+dk)}{\delta B^\alpha(x,k)}
 \right ],
\end{equation}
dropping those terms which do not contribute on account of the antisymmetry of structure 
constant. We calculate infinitesimal Jacobian change as mentioned in \cite{jm} to be
\begin{equation}
\frac{1}{J(k)}\frac{dJ(k)}{dk}= -\int d^4x \left[(\delta A^\alpha _{\mu})\frac{\delta \Theta ^
\prime}{\delta A_\mu ^\alpha}-(\delta c^{\alpha})\frac{\delta \Theta ^\prime}{\delta c^
\alpha} -\delta \bar{c}^\alpha \frac{\delta \Theta ^\prime}{\delta \bar{c}^\alpha}+\delta B^\alpha\frac{\delta \Theta ^\prime}{\delta B^\alpha} \right ],
\label{jc1} 
\end{equation}
Further, it can be shown that the Jacobian in Eq. ({\ref
{jac1}}) can be expressed as 
$e^{iS_1[\phi]}$ if it satisfies the following condition \cite{jm}
\begin{equation}
\int {\cal{D}}\varphi \; e^{i(S_{eff}+S_1)} \left ( \frac{1}{J}\frac{dJ}{d\kappa}-i\frac
{dS_1}{d\kappa}\right )=0 \label{mcond}
\end{equation}
Now, we consider different choices of the parameter $\Theta^\prime$ (which is 
related to $\Theta$ through the relation in Eq. (\ref{80})) to show the connection between a pair of theories. 
\subsection{\large Connecting YM theory in Lorentz gauge to the same theory in axial gauge}
To show the connection between YM theories in Lorentz gauge and axial gauge 
we start with the Lorentz gauge YM theory in the auxiliary field formulation, described by the effective action given in Eq. (\ref{seff}) and
 choose the finite parameter as
\begin{equation} 
\Theta ^\prime =i\int d^4x \; {\bar{c}}^\alpha \left [\gamma_1\lambda B^\alpha + \gamma_2 
\left (\partial \cdot A^\alpha -\eta \cdot A^\alpha \right )\right ],\label{tlab}
\end{equation}
where $\gamma_1$, $\gamma_2$ are arbitrary constants and $\lambda $ is a gauge parameter.
 Using 
Eq. (\ref{jc1}), we calculate the change in the Jacobian of such 
transformation as
\begin{equation}
\frac{1}{J}\frac{dJ}{dk} =i \int d^4x \left [ \gamma_1 \lambda {(B^\alpha)}^2 +\gamma_2 B^
\alpha \left (\partial \cdot A^\alpha -\eta \cdot A^\alpha \right )+\gamma _2 {\bar{c}}^\alpha
 \left( Mc^\alpha -{\tilde {M}}c^\alpha \right )\right ],
\end{equation}
where $ M\equiv\partial\cdot D$ and $\tilde M\equiv\eta\cdot D$.
We further make an ansatz for $S_1$ in this case as
\begin{eqnarray}
S_1\left [\varphi(\kappa), \kappa \right ]&=& \int d^4x\left[\xi_1 (\kappa){(B^\alpha)}^2 +\xi_2 (\kappa ) B^
\alpha \partial\cdot A^\alpha +\xi_3 (\kappa) B^\alpha \eta \cdot A^\alpha \right .\nonumber\\
&+&\left .\xi_4 (\kappa ){\bar{c}}^\alpha Mc^\alpha +\xi_5 (\kappa){\bar{c}}^\alpha{\tilde{M}}c^
\alpha\right]. \label{s1a}
\end{eqnarray}
The constants $\xi_i(\kappa)$  depend on $\kappa$ explicitly and satisfies the 
following initial condition
\begin{equation}
\xi_i(\kappa=0)=0. \label{xcond}
\end{equation}
Using Eq. (\ref{ibr}) we calculate
\begin{eqnarray}
\frac{dS_1}{d\kappa}&=&\int d^4x\left[\frac{d\xi_1}{d\kappa}{(B^\alpha)}^2+\frac{d\xi_2}{d\kappa} B^\alpha \partial \cdot A^
\alpha+\frac{d\xi_3}{d\kappa} B^\alpha \eta\cdot A^\alpha +\frac{d\xi_4}{d\kappa}\bar{c}^\alpha Mc^\alpha +
\frac{d\xi_5}{d\kappa} \bar{c}^\alpha\tilde{M}c^\alpha \right.\nonumber\\
&-&\left.\xi_4 B^\alpha Mc^\alpha \Theta^\prime -\xi_5 B^
\alpha \tilde{M}c^\alpha\Theta^\prime\right].
\end{eqnarray}
From the condition mentioned in Eq. (\ref{mcond}), we obtain
\begin{eqnarray}
\int &{\cal{D}}\varphi & \; \exp\left [i \left (S^L_{eff}+S_1 \right )\right ] \int d^4x \left 
\{ Mc^\alpha\Theta ^\prime[B^\alpha (\xi _2-\xi _4)]+ {\tilde{M}}c^\alpha\Theta^\prime[B^
\alpha(\xi_3-\xi_5)]\right. \nonumber\\
&+&\left. {(B^\alpha)}^2(\frac{d\xi_1}{d\kappa}-\gamma _1\lambda )+B^\alpha\partial\cdot A^\alpha (\frac{d\xi_2}{d\kappa}-\gamma_2) +B^\alpha \eta\cdot A^\alpha ( \frac{d\xi_3}{d\kappa} + \gamma _2)
\right.\nonumber\\
&+&\left.{\bar{c}}^\alpha M c^\alpha (\frac{d\xi_4}{d\kappa}-\gamma_2) +{\bar{c}}^\alpha{\tilde{M}}c^
\alpha (
\frac{d\xi_5}{d\kappa}+\gamma_2 )\right \} = 0. \label{cond}
\end{eqnarray}
The last two terms of Eq. (\ref{cond}) vanishes by using equations of motion for ghost and 
antighost fields when the following condition is satisfied
\begin{equation}
\frac{{d\xi_4}/{d\kappa}-\gamma_2}{\xi_4 -1} =\frac{{d\xi_5}/{d\kappa}+ \gamma_2}{\xi_5}. \label{rc1}
\end{equation}
The non-local $\Theta^\prime$ dependent terms are cancelled by converting them to local terms 
using anti-ghost equation of motion \cite{sdj1}. This can only work if the two $\Theta^\prime$ dependent terms combine in a certain manner, depending again on the ratio of coefficients of $\bar{c}^\alpha 
Mc^\alpha$ and $ \bar{c}^\alpha \tilde{M}c^\alpha$ in terms in $ S^L_{eff}+S_1$. This requires that 
\begin{equation}
\frac{\xi_2 -\xi_4}{\xi_4 -1}=\frac{\xi_3-\xi_5}{\xi_5}.\label{rc2}
\end{equation}
When the above two equations (\ref{rc1}) and (\ref{rc2}) are satisfied, the non-local 
$\Theta^\prime$ dependent terms get converted to local terms. The coefficients of local terms 
${(B^\alpha)}^2$, $ B^\alpha \partial\cdot A^\alpha$, $B^\alpha \eta\cdot A^\alpha$, $\bar{c}^\alpha 
Mc^\alpha$, and $\bar{c}^\alpha \tilde{M}c^\alpha $ independently vanish, giving rise to following 
differential equations respectively,
\begin{eqnarray}
&&\frac{d\xi_1}{d\kappa} -\gamma_1\lambda +\gamma_1\lambda(\xi_2-\xi_4 ) + \gamma_1 \lambda(\xi_3-\xi_5) =0, \nonumber\\
&&\frac{d\xi_2}{d\kappa} -\gamma_2 +\gamma_2 (\xi_2-\xi_4 ) + \gamma_2 (\xi_3-\xi_5) =0, \nonumber\\
&&\frac{d\xi_3}{d\kappa} +\gamma_2 -\gamma_2 (\xi_2-\xi_4 ) - \gamma_2 (\xi_3-\xi_5) =0, \nonumber\\
&&\frac{d\xi_4}{d\kappa} -\gamma_2 =0, \nonumber\\
 &&\frac{d\xi_5}{d\kappa} +\gamma_2 =0. 
\end{eqnarray}
The above equations can be solved for various $\xi_i(\kappa)$ using the boundary conditions 
given by Eq. (\ref{xcond}) and the solutions ($\gamma_2 =1$) are given as
\begin{eqnarray}
\xi_1 &=& \gamma_1\lambda\kappa, \nonumber\\
\xi_2 &=& \kappa, \nonumber\\
\xi_3 &=& -\kappa, \nonumber\\
\xi_4 &=&\kappa, \nonumber\\
\xi_5 &=&-\kappa. \label{sol}
\end{eqnarray}
Putting the above values in Eq. (\ref{s1a}), we get
\begin{equation}
S_1=\gamma_1\lambda\kappa{(B^\alpha)}^2 +\kappa B^\alpha\partial \cdot A^\alpha -\kappa B^
\alpha\eta \cdot A^\alpha +\kappa \bar{c}^\alpha Mc^\alpha -\kappa \bar{c}^\alpha \tilde{M}c^
\alpha. 
\end{equation}
FFBRST in Eq. (\ref{ffb}) with the parameter given in Eq. (\ref{tlab}) connects the 
generating functional in Lorentz gauge,
\begin{equation}
 Z_L=\int {\cal{D}}\varphi e^{iS_{eff}^L}\label{gf}
\end{equation}
 to the
generating functional corresponding to the effective action
\begin{eqnarray}
S^\prime_{eff}&=&S_{eff}^L+S_1(\kappa=1)=  \int d^4x \left[-\frac {1}{4}F_{\mu\nu}^\alpha F^{\alpha\mu\nu} +\frac{\zeta}{2}{(B^\alpha)}^2-B^
\alpha\eta\cdot A^\alpha -\bar{c}^\alpha \tilde{M}c^
\alpha \right],\nonumber\\ 
&=& S_{eff}^{\prime A},
\end{eqnarray}
where $S_{eff}^{\prime A}$ is nothing but FP effective action in axial gauge with the gauge parameter
$\zeta=(2\gamma_1+1)\lambda$.

\subsection{\large Relating theories in Coulomb gauge and Lorentz gauge}
We again  start with Lorentz gauge theory given in Eq. (\ref{seff}) and choose another parameter 
\begin{equation}
\Theta^\prime = i\int d^4x\ \bar{c}^\alpha\left[ \gamma_1\lambda B^\alpha +\gamma_2\left( 
\partial \cdot A^\alpha -\partial^j A_j^\alpha \right ) \right ];\  j=1,2,3, \label{tcg}
\end{equation}
to show the connection with theory in the Coulomb gauge.
The change in the Jacobian due to this FFBRST transformation is calculated using Eq. (\ref{jc1}) as
\begin{equation}
\frac{1}{J}\frac{dJ}{dk} =i \int d^4x \left [ \gamma_1 \lambda {(B^\alpha)}^2 +\gamma_2 B^
\alpha \left (\partial \cdot A^\alpha - \partial^j A_j^\alpha \right )+\gamma _2 {\bar{c}}^\alpha
 \left( Mc^\alpha -{\tilde {M^\prime}}c^\alpha \right )\right ],
\end{equation}
where $\tilde M^\prime =\partial^j D_j$.
We try the following ansatz for $S_1$ for this case as 
\begin{eqnarray}
S_1 &=&\int d^4x \left[ \xi_1 (\kappa ){(B^\alpha)}^2 + \xi_2 (\kappa) B^\alpha \partial \cdot
 A^\alpha + \xi_3 (\kappa )B^\alpha \partial^j A_j^\alpha \right .\nonumber\\
&+&\left .\xi_4 (\kappa) \bar{c}^\alpha Mc^\alpha + \xi_5 (\kappa) \bar{c}^\alpha \tilde M^\prime c^
\alpha \right ] \label{s1cg}
\end{eqnarray}
Now, using the condition Eq. (\ref{mcond}) for replacing the Jacobian as $ e^{iS_1}$ and following the similar 
procedure as discussed in the previous case, we obtain exactly same solutions as given in Eq. (\ref{sol}) for the coefficients $\xi_i$. Putting these solutions in Eq. (\ref{s1cg}) we obtain
\begin{equation}
S_1 =\int d^4x \left [\gamma_1 \lambda \kappa {(B^\alpha)}^2 + \kappa B^\alpha\partial \cdot 
A^\alpha -\kappa B^\alpha\partial^j A^\alpha_j +\kappa{\bar c}^\alpha Mc^\alpha-\kappa{
\bar c}^\alpha\tilde M^\prime c^\alpha\right ].
\end{equation}
The transformed effective action 
\begin{eqnarray}
  S^\prime_{eff} &=&S^L_{eff} +S_1(\kappa =1) \nonumber\\
 &=&\int d^4x \left[-\frac{1}{4}F^{\alpha \mu \nu }F^{\alpha}_{\mu 
\nu}+\frac{\zeta}{2}{(B^\alpha)}^2-B^\alpha\partial ^j A^\alpha _j 
-\bar c^\alpha\tilde M^\prime c^\alpha \right], 
\end{eqnarray}
which is the FP effective action in Coulomb gauge with gauge parameter $\zeta $.

Thus, FFBRST with parameter given in equation (\ref{tcg}) connects the generating functional for YM theories in Lorentz gauge to the generating functional for the same theory in Coulomb gauge.
\subsection{\large FFBRST transformation to link FP effective action in Lorentz gauge to quadratic gauge}
Next we consider theories in quatratic gauges which are often useful in doing
calculations \cite{thoo}.
The effective action in quadratic gauge in terms of auxiliary field can be written as
\begin{eqnarray}
S^Q_{eff} &=&\int d^4x \left[ -\frac{1}{4}F^\alpha_{\mu\nu}F^{\alpha\mu\nu} +\frac{\lambda}{2}{(B^
\alpha)}^2 - B^\alpha \left ( \partial\cdot A^\alpha +d^{\alpha\beta\gamma}A_\mu^\beta A^{
\mu\gamma} \right )\right. \nonumber\\
 &-& \left.\bar{c}^\alpha\partial^\mu {(D_\mu c)}^\alpha  -2d^{\alpha\beta\gamma}\bar{c}^
\alpha A^{\mu\gamma}{(D_\mu c)}^\beta \right ],\label{qg}
\end{eqnarray}
where $d^{\alpha\beta\gamma}$ is structure constant symmetric in $\beta$ and $\gamma$. 
This effective action is invariant under the FFBRST transformation mentioned in Eq. (\ref{ffb}).

For this case, we start with the following choice of the finite field dependent parameter
\begin{equation}
\Theta^\prime = i\int d^4x\ \bar{c}^\alpha \left [ \gamma_1\lambda B^\alpha +\gamma_2 d^{
\alpha\beta\gamma} A_\mu^\beta A^{\mu\gamma} \right ]. \label{qeff}
\end{equation}
We calculate the Jacobian change as
\begin{equation}
\frac{1}{J}\frac{dJ}{d\kappa}=i\int d^4x \left [\gamma_1\lambda {(B^\alpha)}^2 + \gamma_2 B^\alpha 
d^{\alpha\beta\gamma}A_\mu^\beta A^{\mu\gamma} + 2\gamma_2 d^{\alpha\beta\gamma}\bar{c}^\alpha
 (D_\mu c)^\beta A^{\mu\gamma} \right ].
\end{equation}
We make an ansatz for $S_1$ as
\begin{equation}
S_1=\int d^4x \left [\xi_1(\kappa) {(B^\alpha)}^2 +\xi_2 (\kappa) B^\alpha d^{
\alpha\beta\gamma}A_\mu^\beta A^{\mu\gamma} +\xi_3 (\kappa) d^{\alpha\beta\gamma}\bar{c}^
\alpha (D_\mu c)^\beta A^{\mu\gamma} \right ].
\end{equation}
The unknown coefficients $\xi_i$ are determined by using the condition in equation  (\ref{mcond}) and the initial condition in equation (\ref{xcond}), to get
\begin{equation}
S_1(\kappa =1)=\gamma_1\lambda {(B^\alpha )}^2-B^\alpha d^{\alpha\beta\gamma}A^\beta _\mu A^{
\mu\gamma}-2d^{\alpha\beta\gamma}\bar c^\alpha{(D_\mu c)}^\beta A^{\mu\gamma},
\end{equation}
and $ S^\prime_{eff} =S^L_{eff} +S_1(\kappa =1)= S^{\prime Q}_{eff} $, 
which is effective action in quadratic gauge as given in Eq. (\ref{qg}) with gauge parameter $\zeta $.

Thus, the FFBRST with parameter given in equation (\ref{qeff}) connects Lorentz gauge theory to the theory for quadratic gauge.
\subsection{\large FFBRST transformation linking FP action to the most general BRST/anti-BRST invariant action}
The most general BRST/anti-BRST invariant action for YM theories in Lorentz gauge is given as \cite{bath}
\begin{eqnarray}
S_{eff}^{AB}[A,c,\bar{c}] &=& \int d^4x \left[-\frac{1}{4}F^{\alpha \mu \nu }F^{\alpha}_{\mu 
\nu} - \frac{(\partial\cdot A^\alpha)^2}{2\lambda}  + \partial ^\mu\bar{c}^\alpha D_\mu c^\alpha\right. \nonumber 
\\
&+&\left.\frac{\alpha}{2} gf^{\alpha \beta \gamma }\partial\cdot A^\alpha\bar{c}^\beta c^
\gamma -\frac{1}{8}\alpha (1-\frac{1}{2}\alpha )\lambda g^2 f^{\alpha \beta \gamma}\bar{c}^
\beta\bar{c}^\gamma f^{\alpha 
\eta \xi}c^\eta c^\xi \right ].
\end{eqnarray}
This effective action has the following global symmetries.\\
BRST:
\begin{eqnarray}
\delta A_\mu ^\alpha &=&(D_\mu c)^\alpha  \Lambda , \nonumber \\
\delta c^\alpha &=& -\frac{1}{2} g f^{\alpha \beta \gamma }c^\beta c^\gamma \Lambda , 
\nonumber \\
\delta\bar{c}^\alpha &=&\left(\frac{\partial\cdot A^\alpha}{\lambda} -\frac{1}{2}\alpha g f^{
\alpha \beta \gamma} \bar{c}^\beta c^\gamma \right) \Lambda . 
\end{eqnarray}
anti-BRST:
\begin{eqnarray}
\delta A_\mu ^\alpha &=&(D_\mu \bar c)^\alpha \Lambda , \nonumber \\
\delta \bar c^\alpha &=& -\frac{1}{2} g f^{\alpha \beta \gamma }\bar c^\beta \bar c^\gamma \Lambda , 
\nonumber \\
\delta{c}^\alpha &=&\left( -\frac{\partial\cdot A^\alpha}{\lambda} -(1-\frac{1}{2}\alpha) g f^{
\alpha \beta \gamma} \bar{c}^\beta c^\gamma \right) \Lambda .
\end{eqnarray}
The above most general BRST/anti-BRST effective action can be re-expressed in the auxiliary 
field formulation as, 
\begin{eqnarray}
S_{eff}^{AB}[A,c,\bar{c},B] &=& \int d^4x \left[-\frac{1}{4}F^{\alpha \mu \nu }F^{\alpha}_{\mu 
\nu}+\frac{\lambda}{2}(B^\alpha)^2 - B^\alpha(\partial\cdot A^\alpha-\frac{\alpha g \lambda}{2}f^{
\alpha\beta\gamma}\bar{c}^\beta c^\gamma)\right. \nonumber\\
&+&\left. \partial ^\mu\bar{c}D_\mu c -\frac{1}{8}\alpha\lambda g^2f^{\alpha\beta\gamma}\bar{c}^
\beta \bar{c}^\gamma f^{\alpha\eta\xi}c^\eta c^\lambda \right].\label{seffab}
\end{eqnarray}
The off-shell nilpotent, global BRST/anti-BRST symmetries for this effective action are 
given as\\
BRST:
\begin{eqnarray}
\delta A_\mu ^\alpha &=&(D_\mu c)^\alpha \;\Lambda, \nonumber \\
\delta c^\alpha &=& -\frac{1}{2} g f^{\alpha \beta \gamma }c^\beta c^\gamma\; \Lambda, 
\nonumber \\
\delta\bar{c}^\alpha &=& B^\alpha\; \Lambda,\nonumber\\
\delta B^\alpha &=& 0.
\end{eqnarray}
anti-BRST:
\begin{eqnarray}
\delta A^\alpha_\mu &=&{(D_\mu \bar c)}^\alpha \;\Lambda, \nonumber\\
\delta\bar c^\alpha &=&-\frac{1}{2}gf^{\alpha \beta \gamma }\bar c^\beta \bar c^\gamma 
\;\Lambda,\nonumber\\
\delta c^\alpha &=&{(-B^\alpha -gf^{\alpha \beta \gamma}\bar c^\beta c^\gamma )}\;
\Lambda,\nonumber\\
\delta B^\alpha &=&-gf^{\alpha \beta \gamma}B^\beta \bar c^\gamma \;\Lambda.\nonumber\\
\label{antb}
\end{eqnarray}
To obtain the generating functional corresponding to this theory, we apply the FFBRST transformation with the finite field parameter
\begin{equation}
\Theta^\prime =i\int d^4x\ \bar{c}^\alpha\left [ \gamma_1 \lambda  B^\alpha +\gamma_2 f^{\alpha\beta\gamma}\bar{c}^\beta c^\gamma \right ]
\end{equation}
on the generating functional given in Eq. (\ref{gf}).
Using Eq. (\ref{jc1}), change in Jacobian can be calculated as
\begin{equation}
\frac{1}{J}\frac{dJ}{d\kappa}=i\int d^4x \left[ \gamma_1\lambda {(B^\alpha)}^2 +2 \gamma_2 f^{
\alpha\beta\gamma}B^\alpha\bar{c}^\beta c^\gamma -\frac{g}{2} \gamma_2 f^{\alpha \beta\gamma}
\bar{c}^\beta \bar{c}^\gamma f^{\alpha\eta\xi}c^\eta c^\xi \right ].
\end{equation}
We further make an ansatz for $S_1$ as
\begin{equation}
S_1=\int d^4x \left[ \xi_1 (\kappa ) {(B^\alpha)}^2 +\xi_2 (\kappa )B^\alpha f^{\alpha\beta\gamma}\bar{c}^\beta c^
\gamma + \xi_3 (\kappa )f^{\alpha\beta\gamma}\bar{c}^\beta \bar{c}^\gamma f^{\alpha\eta\xi}c^\eta c^
\xi \right ].
\end{equation}
The condition given in equation (\ref{mcond}), 
can be written for this case as
\begin{eqnarray}
&&\int {\cal{D}}\varphi   \; \exp\left [i \left (S^L_{eff}+S_1 \right )\right ] \int d^4x \left 
[ \left (\frac{d\xi_1}{d\kappa} -\gamma_1\lambda \right ){B^\alpha}^2 
+ \left(\frac{d\xi_2}{d\kappa}-2\gamma_2\right) B^\alpha f^{\alpha\beta\gamma}\bar{c}^\beta c^
\gamma \right.  \nonumber \\
&+&\left. \left (\frac{d\xi_3}{d\kappa} +\frac{g}{2}\gamma_2 \right ) f^{\alpha\beta\gamma}
\bar{c}^\beta \bar{c}^\gamma f^{\alpha\eta\xi}c^\eta c^\xi - \left ( \frac{g}{2} \xi_2 
+2 \xi_3 \right )f^{\alpha\beta\gamma} B^\beta\bar{c}^\gamma f^{\alpha \eta\xi}c^\eta c^
\xi \Theta^\prime \right ]=0.\label{mgcond} 
\end{eqnarray}
We look for a special solution corresponding to the condition
\begin{equation}
\frac{g}{2} \xi_2 
+2 \xi_3 =0.\label{sp}
\end{equation}
Comparing the different coefficients, we get the following differential equations for $\xi_i (\kappa )$
\begin{eqnarray}
\frac{d\xi_1}{d\kappa} -\gamma_1\lambda = 0, \\
\frac{d\xi_2}{d\kappa}-2\gamma_2 = 0,\\
\frac{d\xi_3}{d\kappa} +\frac{g}{2}\gamma_2 = 0.
\end{eqnarray}
Solutions of the above equations subjected to the initial condition given in Eq. (\ref{xcond}) are, 
\begin{eqnarray}
\xi_1 &=&\gamma_1 \lambda \kappa,  \nonumber\\
\xi_2 &=& 2\gamma_2 \kappa, \nonumber\\
\xi_3 &=& -\frac{g}{2} \gamma_2\kappa.\label{soln}
\end{eqnarray}
These solutions are consistent with condition in Eq. (\ref{sp}. Since $\gamma_2$ is arbitrary, we choose $\gamma_2=\frac{1}{4} \alpha g \zeta$ 
to get
\begin{eqnarray}
S^L_{eff}+S_1 ( \kappa =1 )&=& \int d^4x \left[-\frac {1}{4}F_{\mu\nu}^\alpha F^{\alpha\mu\nu}+\frac{\zeta}{2}(B^\alpha)^2 
- B^\alpha(\partial\cdot A^\alpha-\frac{\alpha g \zeta}{2}f^{
\alpha\beta\gamma}\bar{c}^\beta c^\gamma)\right. \nonumber\\
&+&\left. \partial ^\mu\bar{c}D_\mu c -\frac{1}{8}\alpha\zeta g^2f^{\alpha\beta\gamma}\bar{c}^
\beta \bar{c}^\gamma f^{\alpha\eta\xi}c^\eta c^\xi \right],\nonumber\\
&=&S^{\prime AB}_{eff}, \label{mga}
\end{eqnarray}
which is the same effective action as mentioned in Eq. (\ref{seffab}), where $\zeta$ is the 
gauge parameter.

Thus even in the auxiliary field formulation the different generating functionals corresponding to the different effective theories can be connect through off-shell nilpotent FFBRST transformation with different choices of the finite parameter which also depends on the auxiliary field. 
\section{\large Finite field dependent anti-BRST formulation}
In this section, we construct the FF anti-BRST transformation analogous to FFBRST transformation. For simplicity, we consider the pure YM theory in Lorentz gauge described by the effective action in Eq. (\ref{sf}) which is invariant under the following on-shell anti-BRST transformation:
\begin{eqnarray}
\delta A_\mu ^\alpha &=&(D_\mu \bar{c})^\alpha\; \Lambda , \nonumber \\
\delta\bar c^\alpha &=& -\frac{1}{2} g f^{\alpha \beta \gamma }\bar{c}^\beta \bar{c}^\gamma 
\;\Lambda , \nonumber \\
\delta{c}^\alpha &=&\left( -\frac{\partial\cdot A^\alpha}{\lambda} - g 
f^{\alpha \beta \gamma} \bar{c}^\beta c^\gamma \right) \;\Lambda, \label{abt} 
\end{eqnarray}
where $\Lambda$ is infinitesimal, anti commuting and global parameter.
Following the procedure similar to the construction of FFBRST as outlined in the 
section II,
we can easily construct the FF anti BRST transformation for the pure YM theory as
\begin{eqnarray}
\delta A_\mu ^\alpha &=&(D_\mu \bar{c})^\alpha \;\Theta , \nonumber \\
\delta\bar c^\alpha &=& -\frac{1}{2} g f^{\alpha \beta \gamma }\bar{c}^\beta \bar{c}^\gamma 
\;\Theta, \nonumber \\
\delta{c}^\alpha &=&\left( -\frac{\partial\cdot A^\alpha}{\lambda} - g 
f^{\alpha \beta \gamma} \bar{c}^\beta c^\gamma \right)\;\Theta, \label{anti}
\end{eqnarray}
where $ \Theta (A,c,\bar c)$ is finite, field dependent and anticommuting parameter.
We would like to investigate the role of such transformation by considering different finite field dependent parameters $ \Theta^\prime(A,c,\bar c)$.
 
\subsection{\large FF anti-BRST transformation to change the gauge parameter }
First we consider a very simple example to outline the procedure. In this example, we show that a simple FF anti-BRST transformation can connect the generating functional corresponding to YM effective action in Lorentz gauge with a gauge parameter $\lambda$ to the generating functional corresponding to same effective action with a different gauge parameter $\lambda^\prime$.
We start with the Lorentz gauge effective action given in Eq. ({\ref {seff}) with the gauge 
parameter $\lambda$ and consider
\begin{equation}
\Theta ^\prime = -i\gamma \int {d^4x\ { c^\alpha (x,\kappa)}\partial\cdot A^\alpha (x,\kappa)},\label{tllb}
\end{equation}
with $\gamma$ as arbitrary parameter. Using the expression given in Eq. (\ref{jc1}), we 
calculate
\begin{equation}
\frac{1}{J}\frac{dJ}{d\kappa}=i\gamma \int d^4x\ \left[\frac{(\partial\cdot A^\alpha )^2}{\lambda} +\bar c^\alpha Mc^\alpha \right].
\end{equation}
We choose
\begin{equation}
S_1=\xi (\kappa )\int d^4x\ \frac{(\partial\cdot A^\alpha )^2}{\lambda}.
\end{equation}
The condition for replacing the Jacobian of the FF anti-BRST transformation with parameter given in 
Eq. (\ref{tllb}) as $e^{iS_1}$ is given in Eq. (\ref{mcond}) and calculated as
\begin{eqnarray}
\int & {\cal{D}}\varphi &  \; \exp\left [i \left (S^L_{eff}+S_1 \right )\right ] \int d^4x \left 
[\frac{(\partial\cdot A^\alpha )^2}{\lambda}(\xi^\prime -\gamma )+2\xi\frac{\partial\cdot A^\alpha}{\lambda}M\bar c^\alpha \Theta^\prime -\bar c^\alpha Mc^\alpha\right]\nonumber\\ 
&=&0.\label{xy}
\end{eqnarray}
The last term of above equation gives no contribution due to dimensional regularization  and we can substitute \cite{jm}
\begin{equation}
\int d^4x\ \frac{\partial\cdot A^\alpha}{\lambda}M\bar c^\alpha \Theta^\prime\rightarrow \gamma\frac{(\partial\cdot A^\alpha )^2}{\lambda}.
\end{equation}  

Thus the LHS of Eq. (\ref{xy}) is vanish iff
\begin{equation}
\xi^\prime -\gamma +2\xi\gamma =0
\end{equation}
We solve this equation subjected to the initial condition given in Eq. (\ref{xcond}) to obtain
\begin{equation}
\xi =\frac{1}{2}(1-e^{-2\gamma\kappa}).
\end{equation}
Thus, at $\kappa=1$ the extra term in the net effective action from the Jacobian is 
\begin{equation}
S_1 =\frac{1}{2}(1-e^{-2\gamma})\int d^4x\ \frac{(\partial\cdot A^\alpha )^2}{\lambda}.
\end{equation}
The new effective action becomes $S^\prime_{eff}=S^L_{eff}+S_1 $. In this case, 
\begin{equation}
S^L_{eff}+S_1=\int d^4x\ \left[-\frac {1}{4}F_{\mu\nu}^\alpha F^{\alpha\mu\nu} -\frac{1}{2\lambda^\prime}{(\partial\cdot A^\alpha)}^2 
-\bar c^\alpha\tilde{M}{c}^\alpha \right] ,
\end{equation}
which is effective action in Lorentz gauge with gauge parameter $\lambda^\prime =\lambda/e^{-2\gamma}$.
Thus, the FF anti-BRST in Eq. (\ref{anti}) with parameter given in Eq. (\ref{tllb}) connects two effective 
theories which differ only by a gauge parameter.

\subsection{\large Lorentz gauge theories to axial gauge theories using FF anti-BRST transformation}
The effective action in Lorentz gauge given in Eq. (\ref{sf}) can be written as 
\begin{equation}
S^L_{eff}= \int d^4x \left[-\frac {1}{4}F_{\mu\nu}^\alpha F^{\alpha\mu\nu} -\frac{1}{2\lambda}({\partial\cdot A^\alpha})^2+ c^\alpha M\bar{c}^\alpha -g f^{
\alpha\beta\gamma}\bar{c}^\beta c^\gamma {(\partial \cdot A)}^\alpha \right],
\end{equation}
where we have interchanged the position of $c, \bar c$ in the ghost term for the seek of convenience.
This action is invariant under anti-BRST transformation given in Eq. (\ref{abt}).
Similarly, the effective action in axial gauge can be written as 
\begin{equation}
S^A_{eff} = \int d^4x \left[-\frac {1}{4}F_{\mu\nu}^\alpha F^{\alpha\mu\nu}-\frac{1}{2\lambda}{(\eta\cdot A^\alpha)}^2 +c^\alpha\tilde{M}\bar{c}^\alpha -g f^{
\alpha\beta\gamma}\bar{c}^\beta c^\gamma {(\eta\cdot A)}^\alpha \right],
\end{equation}
which is invariant under the following anti-BRST symmetry transformations
\begin{eqnarray}
\delta A_\mu ^\alpha &=&(D_\mu \bar{c})^\alpha \;\Lambda , \nonumber \\
\delta c^\alpha &=& -\frac{1}{2} g f^{\alpha \beta \gamma }\bar{c}^\beta \bar{c}^\gamma 
\;\Lambda , 
\nonumber \\
\delta\bar{c}^\alpha &=&\left( -\frac{\eta\cdot A^\alpha}{\lambda} - g 
f^{\alpha \beta \gamma} \bar{c}^\beta c^\gamma \right)\; \Lambda. 
\end{eqnarray}
Now, we show that the generating functionals corresponding to these two effective action are related through FF anti-BRST transformation.

To show the connection, we choose 
\begin{equation}
\Theta^\prime =-i\gamma \int d^4x\ c^\alpha \left (\partial\cdot A^\alpha -\eta \cdot A^\alpha 
\right ).
\end{equation}
We calculate the change in Jacobian corresponding to this FF anti-BRST transformation using Eq. (\ref{jc1}) as
\begin{eqnarray}
\frac{1}{J}\frac{dJ}{d\kappa}&=& i\gamma \int d^4x \left [ \frac{1}{\lambda}{(\partial\cdot A^
\alpha)}^2-\frac{1}{\lambda}(\partial\cdot A^\alpha)(\eta\cdot A^\alpha) +gf^{
\alpha\beta\gamma}\bar{c}^\beta c^\gamma (\partial\cdot A^\alpha)\right. \nonumber \\
&-&\left. gf^{\alpha\beta\gamma}\bar{c}^\beta c^\gamma(\eta\cdot A^\alpha )-c^\alpha M\bar{c}^
\alpha +c^\alpha \tilde
{M}\bar{c}^\alpha \right ].
\end{eqnarray}
We make an ansatz for $S_1$ as the following 
\begin{eqnarray}
S_1&=&\int d^4x\left[\xi_1 (\kappa)(\partial\cdot A^\alpha )^2+\xi_2 (\kappa)(\eta\cdot A^\alpha )^2+\xi_3 (\kappa)(\partial\cdot A^
\alpha )(\eta\cdot A^\alpha )\right] . \nonumber\\
&+&\left .\xi_4 (\kappa)\left (c^\alpha M\bar c^\alpha - gf^{\alpha \beta \gamma}
\bar c^\beta c^\gamma \partial\cdot A^\alpha \right ) 
+\xi_5 (\kappa)\left ( c^\alpha \tilde M\bar c^\alpha - gf^{\alpha \beta \gamma}\bar c^\beta c^\gamma
 \eta\cdot A^\alpha\right )\right],
\end{eqnarray}
where $\xi_i(\kappa)$ are parameters to be determined.
The condition mentioned in Eq. (\ref{mcond} ) to replace the Jacobian as $e^{iS_1} $ for this case is
\begin{eqnarray}
\int &{\cal{D}}\varphi &  \; \exp\left [i \left (S^L_{eff}+S_1 \right )\right ] \int d^4x \left 
[ \left \{ M\bar{c}^\alpha - gf^{\alpha\beta\gamma}\bar{c}^\beta (\partial\cdot A^
\gamma)\right \}\Theta^\prime \left \{ (\partial\cdot A^\alpha)\left( 2\xi_1 +\frac{\xi_4}{
\lambda}\right )\right.\right. \nonumber \\ 
&+&\left. \left.(\eta\cdot A^\alpha)\xi_3\right \}
+ \left \{\tilde{M}\bar{c}^\alpha -gf^{\alpha\beta\gamma}\bar{c}^\beta(\eta\cdot A)^
\gamma\right \} \Theta^\prime \left \{(\partial\cdot A^\alpha)\left( \xi_3 +\frac{\xi_5}{
\lambda}\right) +2\xi_2(\eta\cdot A^\alpha)\right \} \right. \nonumber\\
&+&\left. \left(\frac{d\xi_1}{d\kappa} -\frac{\gamma}{\lambda}\right ){(\partial\cdot A^\alpha)}^2 +
\frac{d\xi_2}{d\kappa} {(\eta \cdot A^\alpha)}^2+(\partial\cdot A^\alpha)(\eta\cdot A^\alpha) 
\left(\frac{d\xi_3}{d\kappa}+\frac{\gamma}{\lambda}\right)\right. \nonumber\\
&+&\left. \left(\frac{d\xi_4}{d\kappa}+\gamma\right) c^\alpha M \bar{c}^\alpha +\left( 
\frac{d\xi_5}{d\kappa}-\gamma \right) c^\alpha\tilde{M}\bar{c}^\alpha - gf^{\alpha\beta\gamma}\bar{c}^
\beta c^\gamma(\partial\cdot A^
\alpha)\left(\frac{d\xi_4}{d\kappa}+\gamma\right) \right. \nonumber\\
&-&\left. g f^{\alpha\beta\gamma}\bar{c}^\beta c^\gamma(\eta\cdot A^\alpha) 
\left(\frac{d\xi_5}{d\kappa}-\gamma\right) \right ]=0. \label{rto}
\end{eqnarray}
The last four terms of Eq. (\ref{rto}) vanish by using equations of motion for ghost and 
antighost field and the non-local $\Theta^\prime$ dependent terms are cancelled by converting them to local terms 
using anti-ghost equation of motion \cite{sdj1}. This occurs only if the two $\Theta^\prime$ dependent terms combine in a certain manner, depending again on the ratio of coefficients of $\bar{c}^\alpha 
Mc^\alpha$ and $ \bar{c}^\alpha \tilde{M}c^\alpha$ in  terms in $ S^L_{eff}+S_1$. i.e. 
\begin{eqnarray}
\frac{2\xi_1 +{\xi_4}/{\lambda}}{\xi_4 +1}&=&\frac{\xi_3 +{\xi_5}/{\lambda}}{\xi_5},\nonumber\\
\frac{\xi_3}{\xi_4 +1}&=&\frac{2\xi_2}{\xi_5}, \nonumber\\
\frac{{d\xi_4}/{d\kappa} +\gamma}{\xi_4 +\gamma}&=&\frac{{d\xi_5}/{d\kappa} -\gamma}{\xi_5}.
\end{eqnarray}
Comparing the coefficients of $(\partial\cdot A^\alpha )^2 $, $(\eta\cdot A^\alpha )^2 $, $(\partial\cdot A^\alpha )(\eta\cdot A^\alpha )$, $ (c^\alpha M \bar{c}^\alpha -gf^{\alpha\beta\gamma}\bar{c}^
\beta c^\gamma\partial\cdot A^
\alpha )$ and $(c^\alpha\tilde{M}\bar{c}^\alpha - g f^{\alpha\beta\gamma}\bar{c}^\beta c^\gamma\eta\cdot A^\alpha)$
\ respectively, we get  
\begin{eqnarray}
&&\frac{d\xi_1}{d\kappa} -\frac{\gamma}{\lambda}+\gamma \left(2\xi_1 +\frac{\xi_4}{\lambda}\right) +\gamma
 \left(\xi_3 +\frac{\xi_5}{\lambda}\right) =0,\label{av} \\
 &&\frac{d\xi_2}{d\kappa} -\gamma \xi_3 -2\gamma \xi_2 =0, \\
 &&\frac{d\xi_3}{d\kappa} +\frac{\gamma}{\lambda}+\gamma \xi_3 -\gamma \left(2\xi_1 +\frac{\xi_4}{\lambda}
\right) +2\gamma \xi_2 -\gamma  \left(\xi_3 +\frac{\xi_5}{\lambda}\right) =0,\\
&&\frac{d\xi_4}{d\kappa}+\gamma =0,\\
&&\frac{d\xi_5}{d\kappa}-\gamma =0. \label{az}
\end{eqnarray}
The solutions of the above equations (\ref{av}) to (\ref{az}) (for $\gamma =1$) are 
\begin{eqnarray}
\xi_1 &=&\frac{1}{2\lambda}\left[1-(\kappa -1)^2\right],\nonumber\\
\xi_2 &=&-\frac{\kappa^2}{2\lambda},\nonumber\\
\xi_3 &=&\frac{1}{\lambda}\kappa (\kappa -1),\nonumber\\
\xi_4 &=&-\kappa ,\nonumber\\
\xi_5 &=&\kappa .
\end{eqnarray}
Putting these in the expression for $S_1$, we have
\begin{eqnarray}
S_1(\kappa=1 )&=&\int d^4x\left[\frac{(\partial\cdot A^\alpha )^2}{2\lambda}-\frac{(\eta\cdot A^\alpha )^2}{2\lambda}-c^
\alpha M\bar c^\alpha +c^\alpha\tilde M\bar c^\alpha +gf^{\alpha \beta \gamma}\bar c^\beta c^
\gamma \partial\cdot A^\alpha \right.\nonumber\\
&-&\left. gf^{\alpha \beta \gamma}\bar c^\beta c^\gamma \eta\cdot A^\alpha\right].
\end{eqnarray}
The new effective action becomes $S^\prime_{eff}=S^L_{eff}+S_1 $. In this case, 
\begin{eqnarray}
S^L+S_1&=&\int d^4x \left[-\frac {1}{4}F_{\mu\nu}^\alpha F^{\alpha\mu\nu} -\frac{1}{2\lambda}{(\eta\cdot A^\alpha)}^2 
+c^\alpha\tilde{M}\bar{c}^\alpha -g f^{\alpha\beta\gamma}\bar{c}^\beta c^\gamma {(\eta\cdot 
A)}^\alpha\right]\nonumber\\
&=& S^A_{eff} 
\end{eqnarray}
which is nothing but the FP effective action in axial gauge.

Thus, the generating functional corresponding to Lorentz gauge and axial gauge can also be related by FF anti-BRST transformation. We observe FF anti-BRST transformation plays exactly the same role as FFBRST transformation in this example.
\subsection{\large Connection of YM theories in Lorentz gauge to same theories in Coulomb gauge through FF anti-BRST transformation}
To show the connection between generating functional corresponding to the effective action in Lorentz gauge to that of the effective action in Coulomb gauge through FF anti BRST transformation, we choose the parameter, 
\begin{equation}
\Theta^\prime =-i\gamma \int d^4x\ c^\alpha (\partial\cdot A^\alpha -\partial_j A^{j\alpha})
\end{equation}
Using equation (\ref{jc1}), we calculate the change in Jacobian as
\begin{eqnarray}
{\frac{1}{J}\frac{dJ}{d\kappa}}&=&i\gamma \int d^4x\left[\frac{(\partial\cdot A^\alpha )^2}{
\lambda}-\frac{(\partial\cdot A^\alpha )(\partial_j A^{j\alpha})}{\lambda}- c^\alpha M\bar c^
\alpha + c^\alpha\tilde M^\prime \bar c^\alpha +gf^{\alpha \beta \gamma}\bar c^\beta c^\gamma 
\partial\cdot A^\alpha \right.\nonumber\\
&-&\left. gf^{\alpha \beta \gamma}\bar c^\beta c^\gamma \partial_j A^{j\alpha} \;\right]
\end{eqnarray}
where $\tilde M^\prime =\partial_jD^j$.
We make an ansatz for $S_1$ looking at the different terms in the effective action in Lorentz gauge and Coulomb gauge as
\begin{eqnarray}
S_1&=&\int d^4x\left[\xi_1 (\kappa)(\partial\cdot A^\alpha )^2+\xi_2 (\kappa)(\partial_j A^{j\alpha})^2+\xi_3 (\kappa)(\partial\cdot A^
\alpha )(\partial_j A^{j\alpha})\right. \nonumber\\
&+&\left.\xi_4 (\kappa )\left (c^\alpha M\bar c^\alpha
- gf^{\alpha \beta \gamma}\bar c^\beta c^\gamma \partial\cdot A^\alpha\right )
+\xi_5 (\kappa)\left ( c^\alpha \tilde M^\prime\bar c^\alpha  - gf^{\alpha \beta \gamma}\bar c^\beta c^\gamma \partial_j A^{j\alpha}\right )\right]
\end{eqnarray}
$S_1$ will be the part of the new effective action if and only if the condition in Eq. (\ref{mcond}) is satisfied. The 
condition in this particular case reads as
\begin{eqnarray}
\int &{\cal{D}}\varphi &  \; \exp\left [i \left (S^L_{eff}+S_1 \right )\right ] \int d^4x \left 
[  \left\{ M\bar{c}^\alpha - gf^{\alpha\beta\gamma}\bar{c}^\beta (\partial\cdot A^
\gamma)\right \}\Theta^\prime \left \{ (\partial\cdot A^\alpha)\left( 2\xi_1 +\frac{\xi_4}{
\lambda}\right )\right.\right. \nonumber \\ 
&+&\left. \left.(\partial_j A^{j\alpha})\xi_3\right \}
+ \left \{\tilde{M}^\prime\bar{c}^\alpha -gf^{\alpha\beta\gamma}\bar{c}^\beta(\partial_j {A^j}^
\gamma)\right \} \Theta^\prime \left \{(\partial\cdot A^\alpha)\left( \xi_3 +\frac{\xi_5}{
\lambda}\right) +2\xi_2(\partial_j A^{j\alpha})\right \} \right. \nonumber\\
&+&\left. \left( \frac{d\xi_1}{d\kappa}-\frac{\gamma}{\lambda}\right ){(\partial\cdot A^\alpha)}^2 +
\frac{d\xi_2}{d\kappa}{(\partial_j A^{j\alpha})}^2+(\partial\cdot A^\alpha)(\partial_j A^{j\alpha}) 
\left(\frac{d\xi_3}{d\kappa}+\frac{\gamma}{\lambda}\right)\right. \nonumber\\
&+&\left. \left(\frac{d\xi_4}{d\kappa}+\gamma\right) c^\alpha M \bar{c}^\alpha +\left( 
\frac{d\xi_5}{d\kappa}-\gamma \right)c^\alpha\tilde{M}^\prime\bar{c}^\alpha - gf^{\alpha\beta\gamma}\bar{c}^
\beta c^\gamma (\partial\cdot A^\alpha )\left (\frac{d\xi_4}{d\kappa}+\gamma\right ) \right. \nonumber\\
&-&\left. g f^{\alpha\beta\gamma}\bar{c}^\beta c^\gamma(\partial_j A^{j\alpha}) 
\left(\frac{d\xi_5}{d\kappa}-\gamma\right) \right ]=0. 
\end{eqnarray}
Following the similar procedure as in subsection IV(A), we obtain $S_1$ at $\kappa =1$ as
\begin{eqnarray}
S_1&=&\int d^4x\left[\frac{(\partial\cdot A^\alpha )^2}{2\lambda}-\frac{(\partial_j A^{j\alpha})^2}{2\lambda
}-c^\alpha M\bar c^\alpha +c^\alpha\tilde M\bar c^\alpha +gf^{\alpha \beta \gamma}\bar c^\beta
 c^\gamma \partial\cdot A^\alpha \right.\nonumber\\
&-&\left. gf^{\alpha \beta \gamma}\bar c^\beta c^\gamma \partial_j 
A^{j\alpha}\right].
\end{eqnarray}
Adding this part to $S_{eff}^L$ we obtain 
\begin{eqnarray}
S^L_{eff}+S_1(\kappa=1) &=&\int d^4x \left[-\frac{1}{4}F^{\alpha \mu \nu }F^{\alpha}_{\mu 
\nu}-\frac{(\partial_j A^{j\alpha})^2}{2\lambda}+c^\alpha\tilde M^\prime\bar 
c^\alpha -gf^{\alpha \beta \gamma}\bar c^\beta c^\gamma \partial_j A^{j\alpha}\right]\nonumber\\
&=&\int d^4x \left[-\frac{1}{4}F^{\alpha \mu \nu }F^{\alpha}_{\mu 
\nu}-\frac{(\partial_j A^{j\alpha})^2}{2\lambda}-\bar c^\alpha\tilde M^\prime 
c^\alpha\right]\nonumber\\
&=& S^C_{eff},
\end{eqnarray}

which is effective action in Coulomb gauge.
Thus, the generating functionals corresponding to Lorentz gauge and Coulomb gauge can also be related by FF anti-BRST transformation.

\subsection{\large FF anti-BRST transformation connecting FP action to most general BRST/anti-BRST invariant action }
In order to connect these two theories viz. YM effective action in Lorentz gauge and the most 
general BRST/anti-BRST invariant action in Lorentz gauge, we consider finite field dependent parameter as
\begin{equation}
\Theta^\prime =-i\gamma\int d^4x\ c^\alpha f^{\alpha \beta \gamma}\bar c^\beta c^\gamma.
\end{equation}
Then, corresponding to the above $\Theta^\prime$ the change in Jacobian, using the Eq.(\ref{jc1}), is 
calculated as
\begin{equation}
\frac{1}{J}\frac{dJ}{d\kappa}=i\gamma\int d^4x\left[2\frac{\partial\cdot A^\alpha}{\lambda}f^{
\alpha \beta \gamma}\bar c^\beta c^\gamma +gf^{\alpha \beta \gamma}\bar c^\beta c^\gamma f^{
\alpha\eta\xi}\bar c^\eta c^\xi \right]
\end{equation}
Looking at the kind of terms present in the FP effective action in Lorentz gauge 
and in the most general BRST/anti-BRST invariant effective action in Lorentz gauge, we try an ansatz for $S_1$ as
\begin{equation}
S_1=\int d^4x\left[\xi_1 (\kappa)f^{\alpha \beta \gamma}\partial\cdot A^\alpha \bar c^\beta c^\gamma +\xi_2 (\kappa) f^{\alpha
 \beta \gamma}\bar c^\beta \bar c^\gamma f^{\alpha\eta\xi}c^\eta c^\xi\right].
\end{equation}
$S_1$ can be expressed as $e^{iS_1}$ iff it satisfies the condition mentioned in Eq. (\ref
{mcond}). The condition for this case is calculated as 
\begin{eqnarray}
\int &{\cal{D}}\varphi &  \; \exp\left [i \left (S^L_{eff}+S_1 \right )\right ] \int d^4x\left
[f^{\alpha \beta \gamma }\partial\cdot A^\alpha\bar c^\beta c^\gamma \left (\frac{d\xi_1}{d\kappa} -
\frac{2\gamma}{\lambda}\right )\right.\nonumber\\
&+& f^{\alpha \beta \gamma}\bar c^\beta \bar c^\gamma f^{\alpha\eta\xi}c^\eta c^\xi \left (
\frac{d\xi_2}{d\kappa} +\frac{\gamma g}{2} -\gamma\xi_1 \right)\nonumber\\
&+& \left. f^{\alpha \beta \gamma}\bar c^\beta \bar c^\gamma f^{\alpha\eta\xi} c^
\eta\partial\cdot A^\xi \Theta^\prime\left(\frac{\xi_1^2}{2} -\frac{g \xi_1}{2}-\frac{2
\xi_2}{\lambda}\right )\right ]=0.
\end{eqnarray}
We look for a special solution  corresponding to the condition 
\begin{equation}
\frac{\xi_1^2}{2} -\frac{g \xi_1}{2}-\frac{2\xi_2}{\lambda}=0.
\end{equation}
The coefficient of $ f^{\alpha \beta \gamma}\partial\cdot A^\alpha \bar c^\beta c^\gamma $ and
$ f^{\alpha \beta \gamma}\bar c^\beta \bar c^\gamma f^{\alpha\eta\xi}c^\eta c^\xi $ gives 
respectively
\begin{equation}
\frac{d\xi_1}{d\kappa} -\frac{2\gamma}{\lambda}=0,
\end{equation}
\begin{equation}
\frac{d\xi_2}{d\kappa} +\frac{g\gamma}{2}-\gamma\xi_1 =0.
\end{equation}
For a particular $\gamma =\frac{\alpha\lambda g}{4}$,
the solutions of above two equations are 
\begin{equation}
\xi_1 =\frac{\alpha}{2}g\kappa, 
\end{equation}
\begin{equation}
\xi_2 =-\frac{\alpha}{8}\lambda g^2\kappa +\frac{\alpha^2}{16}\lambda g^2\kappa^2.
\end{equation}
At $\kappa=1$
\begin{equation}
S_1=\int d^4x\left[\frac{\alpha}{2}gf^{\alpha \beta \gamma}\partial\cdot A^\alpha \bar c^\beta c^\gamma -\frac
{\alpha}{8}\left(1-\frac{\alpha}{2}\right)\lambda  g^2f^{\alpha \beta \gamma}\bar c^\beta \bar c^\gamma 
f^{\alpha\eta\xi}c^\eta c^\xi\right].
\end{equation}
Hence,
\begin{eqnarray}
S^L_{eff}+S_1&=& \int d^4x \left[-\frac{1}{4}F^{\alpha \mu \nu }F^{\alpha}_{\mu 
\nu} - \frac{(\partial\cdot A^\alpha)^2}{2\lambda }  + \partial ^\mu\bar{c}D_\mu c\right. \nonumber 
\\
&+&\left.\frac{\alpha}{2} gf^{\alpha \beta \gamma }\partial\cdot A^\alpha\bar{c}^\beta c^
\gamma -\frac{1}{8}\alpha (1-\frac{1}{2}\alpha )\lambda g^2 f^{\alpha \beta \gamma}\bar{c}^
\beta\bar{c}^\gamma f^{\alpha 
\eta \xi}c^\eta c^\xi \right ]\nonumber\\
&=& S_{eff}^{AB}[A,c,\bar{c}].
\end{eqnarray}
which is most general BRST/anti-BRST invariant effective action.
 
Thus, the generating functional corresponding to most general effective action in Lorentz gauge can also be related through FF anti-BRST transformation.
\section{\large OFF-SHELL NILPOTENT FF anti-BRST }
The FF anti-BRST transformation, we have constructed in previous section, is on-shell nilpotent. In this section, we construct FF anti- BRST transformation which is off-shell nilpotent. For this purpose, we consider the following effective action for YM theories in auxiliary field formulation in Lorentz gauge 
\begin{equation}
S^L_{eff}=\int d^4x \left[-\frac{1}{4}F^{\alpha \mu \nu }F^{\alpha}_{\mu 
\nu}+\frac{\lambda}{2}{(B^\alpha)}^2-B^\alpha\partial\cdot A^\alpha +c^\alpha M\bar 
c^\alpha - gf^{\alpha\beta\gamma}\bar c^\beta c^\gamma {(\partial\cdot A)}^\alpha \right]. \label{act}
\end{equation}
This effective action is invariant under anti-BRST transformation mentioned in Eq. (\ref{antb}).
Following the procedure outlined in Sec. III, we obtain the FF anti-BRST transformation in auxiliary field formulation as,
\begin{eqnarray}
\delta A^\alpha_\mu &=&{(D_\mu \bar c)}^\alpha \ \Theta  (A,c,\bar c,B),\nonumber\\
\delta\bar c^\alpha &=&-\frac{1}{2}gf^{\alpha \beta \gamma }\bar c^\beta \bar c^\gamma \ \Theta
  (A,c,\bar c,B), \nonumber\\
\delta c^\alpha &=&{(-B^\alpha -gf^{\alpha \beta \gamma}\bar c^\beta c^\gamma )}\ \Theta  (A,c,
\bar c,B), \nonumber\\
\delta B^\alpha &=&-gf^{\alpha \beta \gamma}B^\beta \bar c^\gamma\ \Theta  (A,c,\bar c,B),
\end{eqnarray}
which also
leaves the effective action in Eq. (\ref{act}) invariant.
Now, we consider the different choices of the parameter $\Theta^\prime (A,c,\bar c,B)$ in auxiliary 
field formulation to connect different theories.
\subsection{\large FF anti-BRST transformation connecting YM theories in Lorentz gauge to Coulomb gauge}
The effective action for YM theory in Coulomb gauge can be written after rearranging the ghost term as
\begin{equation}
S^C_{eff}=\int d^4x \left[-\frac{1}{4}F^{\alpha \mu \nu }F^{\alpha}_{\mu 
\nu}+\frac{\lambda}{2}{(B^\alpha)}^2-B^\alpha\partial ^j A^\alpha _j +c^\alpha\tilde 
M^\prime\bar c^\alpha - gf^{\alpha\beta\gamma}\bar c^\beta c^\gamma \partial _j A^{j\alpha }\right].
\end{equation}
To show the connection of this theory with the theory in Lorentz gauge, we choose the finite field dependent parameter 
\begin{equation}
\Theta ^\prime =-i\int d^4x\ c^\alpha[\gamma _1\lambda B^\alpha +\gamma _2(\partial\cdot A^
\alpha -\partial ^jA_j^\alpha )].
\end{equation}
We note that this parameter is different from the parameter of the FFBRST in Section III B.
Using the above $\Theta^\prime$, we find the change in Jacobian as
\begin{eqnarray}
\frac {1}{J}\frac{dJ}{d\kappa}&=&i\int d^4x\left[\lambda\gamma_1 (B^\alpha )^2+\gamma_2 B^
\alpha\partial\cdot A^\alpha -\gamma_2 B^\alpha \partial^j A_j^\alpha +\gamma_2 gf^{\alpha 
\beta \gamma}\bar c^\beta c^\gamma\partial\cdot A^\alpha \right.\nonumber\\
&-&\left .\gamma_2 gf^{\alpha \beta \gamma}\bar c^\beta c^\gamma\partial^j A^\alpha _j -
\gamma_2 cM\bar c+\gamma_2 c\tilde M^\prime\bar c\right ].
\end{eqnarray}
We make an ansatz for $S_1$ as
\begin{eqnarray}
&&S_1=\int d^4x\left[\xi _1 (\kappa)(B^\alpha )^2+\xi _2 (\kappa)B^\alpha\partial\cdot A^\alpha+\xi_3 (\kappa) B^\alpha\partial ^jA_j^
\alpha \right. \nonumber\\
&+&\left. \xi _4 (\kappa )\left (c^\alpha M\bar c^\alpha
- gf^{\alpha \beta \gamma}\bar c^\beta c^\gamma
\partial\cdot A^\alpha \right )
+\xi_5 (\kappa)\left (c^\alpha\tilde M^\prime\bar c^\alpha -gf^{\alpha\beta\gamma}\bar c^\beta c^\gamma \partial _j A^{j\alpha}\right )\right].
\end{eqnarray}
The essential requirement for replacing the Jacobian as $e^{iS_1}$ mentioned in Eq. (\ref{mcond}) is satisfied iff
\begin{eqnarray}
\int &{\cal{D}}\varphi &  \; \exp\left [i \left (S^L_{eff}+S_1 \right )\right ]\int d^4x\left
\{ (B^\alpha )^2(\frac{d\xi_1}{d\kappa}-\lambda\gamma_1 )+B^\alpha\partial\cdot A^\alpha (\frac{d\xi_2}{d\kappa}-\gamma_2 )\right.\nonumber\\
&+&\left. B^\alpha\partial^j A_j^\alpha (\frac{d\xi_3}{d\kappa}+\gamma_2 )
-\bar c^\alpha Mc^\alpha (\frac{d\xi_4}{d\kappa}+\gamma_2 )-\bar c^\alpha \tilde Mc^\alpha (
\frac{d\xi_5}{d\kappa}-\gamma_2 )+[M\bar c^\alpha  \right.\nonumber\\
&-&\left .gf^{\alpha\beta\gamma}\bar c^\beta\partial\cdot 
A^\gamma ]\Theta^\prime[B^\alpha (\xi_2 +\xi_4 )]
+[\tilde M^\prime\bar c^\alpha -gf^{\alpha\beta\gamma}\bar c^\beta\partial^j A_j^\gamma ]
\Theta^\prime[B^\alpha (\xi_3 +\xi_5 )]\right\}\nonumber\\
&=&0 \label{ac}
\end{eqnarray}
In the above equation the two ghost terms vanish using equations of motion for ghost and antighost 
fields and nonlocal terms become local, only if, it satisfy the following conditions
\begin{equation}
 \frac{{d\xi_4}/{d\kappa}+\gamma_2}{\xi_4 +1}=\frac{{d\xi_5}/{d\kappa}-\gamma _2}{\xi_5},
\end{equation}
\begin{equation}
\frac{\xi_2 +\xi_4}{\xi_4 +1}=\frac{\xi_3 +\xi_5}{\xi_5}. 
\end{equation}
We further obtain equations for the parameter $\xi_i $ by vanishing the coefficient of different independent terms in the LHS of the Eq. (\ref{ac}) as 
\begin{eqnarray}
&&\frac{d\xi_1}{d\kappa} -\gamma_1\lambda +\gamma_1\lambda(\xi_2 +\xi_4 ) + \gamma_1 \lambda(\xi_3 +\xi_5) =0, \nonumber\\
&&\frac{d\xi_2}{d\kappa} -\gamma_2 +\gamma_2 (\xi_2 +\xi_4 ) + \gamma_2 (\xi_3 +\xi_5) =0, \nonumber\\
&&\frac{d\xi_3}{d\kappa} +\gamma_2 -\gamma_2 (\xi_2 +\xi_4 ) - \gamma_2 (\xi_3 +\xi_5) =0, \nonumber\\
&&\frac{d\xi_4}{d\kappa}+\gamma_2 =0, \nonumber\\
&& \frac{d\xi_5}{d\kappa} -\gamma_2 =0. 
\end{eqnarray}
We determine the parameter $\xi_i$ subjected to the initial condition in Eq. (\ref{xcond}) as 
\begin{eqnarray}
\xi_1 &=& \gamma_1\lambda\kappa, \nonumber\\
\xi_2 &=& \kappa, \nonumber\\
\xi_3 &=& -\kappa, \nonumber\\
\xi_4 &=& -\kappa, \nonumber\\
\xi_5 &=& \kappa.
\end{eqnarray}

Using the above solutions for $\xi_i$, we write  $S_1$ at $\kappa =1$ as
\begin{eqnarray}
S_1&=&\int d^4x\left[\lambda\gamma_1 (B^\alpha )^2 +B^\alpha\partial\cdot A^\alpha -B^\alpha\partial_j A^{j
\alpha} -c^\alpha M\bar c^\alpha +gf^{\alpha\beta\gamma}\bar c^\alpha c^\beta\partial\cdot A^
\gamma +c^\alpha\tilde M^\prime\bar c^\alpha \right.\nonumber\\
&-&\left. gf^{\alpha\beta\gamma}\bar c^\beta c^\gamma\partial_j A^{j\alpha}\right].
\end{eqnarray}
Now, when this $S_1$ is added to the effective action $S^L_{eff}$, it provides effective action in Coulomb gauge as
\begin{eqnarray} 
S^L_{eff} +S_1&=&\int d^4x \left[-\frac{1}{4}F^{\alpha \mu \nu }F^{\alpha}_{\mu 
\nu}+\frac{\zeta}{2}{(B^\alpha)}^2-B^\alpha\partial ^j A^\alpha _j 
+c^\alpha\tilde M^\prime\bar c^\alpha - gf^{\alpha\beta\gamma}\bar c^\beta c^\gamma \partial _j 
A^{j\alpha}\right], \nonumber\\
&=&\int d^4x \left[-\frac{1}{4}F^{\alpha \mu \nu }F^{\alpha}_{\mu 
\nu}+\frac{\zeta}{2}{(B^\alpha)}^2-B^\alpha\partial ^j A^\alpha _j 
-\bar c^\alpha\tilde M^\prime c^\alpha \right],\nonumber\\
&=& S^{\prime C}_{eff},
\end{eqnarray}
which is the effective action in Coulomb gauge with gauge parameter $\zeta$. 
 
 Thus, FF anti-BRST in auxiliary field formulation produce the same result as expected, even though the finite finite parameter is different. 
\subsection {\large FF anti-BRST transformation connecting YM theories in Lorentz gauge to axial gauge}
We repeat the same steps as in the previous subsection again for this case, with a different finite field dependent parameter  
\begin {equation}
\Theta^\prime =-i\int d^4x c^\alpha\left[\gamma_1\lambda B^\alpha +\gamma_2 (\partial\cdot A^
\alpha -\eta\cdot A^\alpha )\right ],\label{the}
\end{equation}
and consider the ansatz for $S_1$ as
\begin{eqnarray}
S_1&=&\int d^4x\left[\xi_1 (\kappa)(B^\alpha )^2 +\xi_2 (\kappa)B^\alpha\partial\cdot A^\alpha +\xi_3 (\kappa)B^\alpha\eta\cdot A^
\alpha +\xi_4 (\kappa) (c^\alpha M\bar c^\alpha \right.\nonumber\\ 
&-& gf^{\alpha 
\beta \gamma}\bar c^\beta c^\gamma \partial\cdot A^\alpha )
+\left.\xi_5 (\kappa )( c^\alpha\tilde M\bar c^\alpha
- gf^{\alpha \beta \gamma}\bar c^\beta c^\gamma \eta\cdot A^\alpha )\right].
\end{eqnarray}
The condition for which the Jacobian of FF anti-BRST transformation in auxiliary field formulation corresponding to the parameter given in Eq. (\ref{the}) can be replaced as $e^{iS_1}$ is,
\begin{eqnarray}
\int &{\cal{D}}\varphi &  \; \exp\left [i \left (S^L_{eff}+S_1 \right )\right ] \int d^4x\left
[ \left \{ M\bar{c}^\alpha - gf^{\alpha\beta\gamma}\bar{c}^\beta (\partial\cdot A^
\gamma)\right \}\Theta^\prime \left \{ B^\alpha (\xi_2 +\xi_4 )\right \}\right. \nonumber\\
&+&\left.\left \{ \tilde M\bar{c}^\alpha - gf^{\alpha\beta\gamma}\bar{c}^\beta (\eta\cdot A^
\gamma)\right \}\Theta^\prime \left \{ B^\alpha (\xi_3+\xi_5 )\right \} +(B^\alpha )^2 (
\frac{d\xi_1}{d\kappa} -\lambda\gamma_1 )\right.\nonumber\\
&+&\left. B^\alpha\partial\cdot A^\alpha (\frac{d\xi_2}{d\kappa} -\gamma_2 ) + B^\alpha\eta\cdot A^
\alpha(\frac{d\xi_3}{d\kappa} +\gamma_2 )+c^\alpha M\bar c^\alpha (\frac{d\xi_4}{d\kappa} +\gamma_2 )
\right.\nonumber\\
&+&\left. c^\alpha \tilde M\bar c^\alpha (\frac{d\xi_5}{d\kappa} -\gamma_2 )-gf^{\alpha \beta \gamma}
\bar c^\beta c^\gamma \partial\cdot A^\alpha (\frac{d\xi_4}{d\kappa} +\gamma_2 )\right.\nonumber\\
&-&\left. gf^{\alpha \beta \gamma}\bar c^\beta c^\gamma \eta\cdot A^\alpha (\frac{d\xi_5}{d\kappa} -
\gamma_2 )\right] =0.
\end{eqnarray}
The parameters $\xi_i$ are determined using this condition and we obtain the extra piece of the action $S_1$ as
\begin{eqnarray}
S_1&=&\int d^4x\left[\lambda\gamma_1 ( B^\alpha )^2 + B^\alpha\partial\cdot A^\alpha - B^\alpha\eta\cdot A^
\alpha - c^\alpha M\bar c^\alpha+ c^\alpha\tilde M\bar c^\alpha + gf^{\alpha \beta \gamma}\bar 
c^\beta c^\gamma \partial\cdot A^\alpha\right. \nonumber\\
&-&\left.gf^{\alpha \beta \gamma}\bar c^\beta c^\gamma \eta\cdot A^\alpha\right].
\end{eqnarray}
Now,
\begin{eqnarray}
 S^L_{eff}+S_1&=& \int d^4x \left[-\frac{1}{4}F^{\alpha \mu \nu }F^{\alpha}_{\mu 
\nu} +\frac{\zeta}{2} (B^\alpha )^2 - B^\alpha\eta\cdot A^\alpha + 
c^\alpha\tilde M\bar c^\alpha -gf^{\alpha \beta \gamma}\bar c^\beta c^\gamma \eta\cdot 
A^\alpha\right]\nonumber\\
&=&S^{\prime A}_{eff},
\end{eqnarray}
 where $S^{\prime A}_{eff}$ is the effective action in axial gauge
with gauge parameter $\zeta$.
 
 This implies off-shell nilpotent FF anti-BRST also produces the same result even though calculations are different.
\subsection{\large FF anti-BRST transformation to connect most general BRST/anti-BRST invariant action }
We consider one more example in FF anti-BRST formulation using auxiliary field. We show that the most general BRST/anti-BRST invariant theory can be obtained from FP theory in Lorentz gauge. 
The most general effective action which is invariant under BRST/anti-BRST transformation in auxiliary field is given in Eq. (\ref{seffab}). 

We choose
\begin{equation}
\Theta^\prime =-i\int d^4x\ {c}^\alpha\left [ \gamma_1 \lambda  B^\alpha +\gamma_2 f^{\alpha\beta\gamma}\bar{c}^\beta c^\gamma \right ].
\end{equation} 

and take the ansatz for $S_1$ as
\begin{equation}
S_1=\int d^4x\left[\xi_1 (\kappa )(B^\alpha)^2 +\xi_2 (\kappa)B^\alpha f^{\alpha\beta\gamma}\bar c^\beta c^\gamma +\xi_3 (\kappa)f^{\alpha\beta\gamma}\bar
 c^\beta \bar c^\gamma f^{\alpha\eta\xi}c^\eta c^\xi\right].
\end{equation}
The condition in Eq. (\ref{mcond}) leads to 
\begin{eqnarray}
&&\int {\cal{D}}\varphi   \; \exp\left [i \left (S^L_{eff}+S_1 \right )\right ] \int d^4x \left 
[ \left (\frac{d\xi_1}{d\kappa} -\gamma_1\lambda \right ){B^\alpha}^2 
+ \left(\frac{d\xi_2}{d\kappa}-2\gamma_2\right) B^\alpha f^{\alpha\beta\gamma}\bar{c}^\beta c^
\gamma \right.  \nonumber \\
&+&\left. \left (\frac{d\xi_3}{d\kappa} +\frac{g}{2}\gamma_2 \right ) f^{\alpha\beta\gamma}
\bar{c}^\beta \bar{c}^\gamma f^{\alpha\eta\xi}c^\eta c^\xi - \left ( \frac{g}{2} \xi_2 
+2 \xi_3 \right )f^{\alpha\beta\gamma} B^\beta{c}^\gamma f^{\alpha \eta\xi}\bar c^\eta \bar c^
\xi \Theta^\prime \right ]=0.
\end{eqnarray}
Using the same procedure we obtain the solutions for the parameter $\xi_i$, exactly same as in Eq. (\ref{soln}).
Even if the finite parameter is different in FFBRST and FF anti-BRST, we obtain the same contribution from Jacobian in this case as
\begin{equation}
S_1=\int d^4x\left[\gamma_1\lambda (B^\alpha )^2+\frac{\alpha g\zeta}{2}B^\alpha f^{\alpha\beta\gamma}\bar{c}^\beta c^\gamma -\frac{1}{8}
\alpha g^2\zeta f^{\alpha\beta\gamma}\bar{c}^\beta\bar{c}^\gamma f^{\alpha\eta\xi}c^\eta c^
\xi\right].
\end{equation}
Now,
\begin{equation}
S^L_{eff}+S_1 =S_{eff}^{\prime AB},
\end{equation}
which is nothing but most general BRST/anti-BRST invariant effective action with gauge parameter $\zeta$ mentioned in Eq. (\ref{mga}).
In all three cases we show that off-shell nilpotent FF anti-BRST plays exactly same role as on-shell nilpotent FF anti-BRST.
\section{\large conclusion}
In this work we have started with the reformulation of the FFBRST transformation in an auxiliary field formulation where 
the BRST transformation is off-shell nilpotent. We have considered several examples with different choices of finite parameter to connect the different effective theories. In auxiliary field formulation the finite parameters depend also on B field hence different from FFBRST formulation without B field. However, most of the results of the FFBRST transformation 
are also obtained in auxiliary field formulation. In this paper we have introduced and developed for first time the concept of the FF 
anti-BRST 
transformation analogous to the FFBRST transformation. FF anti-BRST transformation can also be 
used to connect the different generating functionals corresponding to different effective theories. 
Several examples have been worked out explicitly to show the results. Lastly we consider the 
FF anti-BRST transformation also in auxiliary field formulation to make it off-shell 
nilpotent. The overall multiplicative antighost field 
in the finite parameters of the FFBRST transformations is replaced by ghost field in case of the FF anti-BRST transformation. Even though the finite parameters and hence the calculations are different, the same results are also produced in 
an auxiliary field formulation of the FF anti-BRST transformation. The BRST and the anti-BRST 
transformations are not independent transformations in the YM theories. We observe that the FF 
anti-BRST transformations play exactly the same role in connecting theories in 1-form gauge 
theory as expected. In 2-form gauge theories the BRST and the anti-BRST transformations play some sort of independent roles. Therefore, it will be interesting to study the finite field dependent BRST 
and anti-BRST transformations in 2-form gauge theories \cite{sm}. \\
\\
\noindent
{\Large{\bf {Acknowledgment}}} \\

\noindent
We thankfully acknowledge the financial support
from the Department of Science and Technology (DST), Government of India, under
the SERC project sanction grant No. SR/S2/HEP-29/2007.\\

\end{document}